\documentclass[10pt,conference]{IEEEtran}
\IEEEoverridecommandlockouts

\usepackage{cite}
\usepackage{amsmath,amssymb,amsfonts}
\usepackage{algorithmic}
\usepackage{pgfplots}
\pgfplotsset{compat=1.17}
\usepackage{graphicx}
\usepackage{textcomp}
\usepackage{xcolor}
\definecolor{darkred}{rgb}{0.55, 0, 0}
\definecolor{darkblue}{rgb}{0, 0, 0.55}
\definecolor{darkgreen}{rgb}{0, 0.39, 0}
\usepackage{url}
\def\BibTeX{{\rm B\kern-.05em{\sc i\kern-.025em b}\kern-.08em
    T\kern-.1667em\lower.7ex\hbox{E}\kern-.125emX}}
    
\begin{document}
\title{CI at Scale: Lean, Green, and Fast}

\author{\IEEEauthorblockN{Dhruva Juloori,
Zhongpeng Lin,
Matthew Williams, 
Eddy Shin,
Sonal Mahajan}
\IEEEauthorblockA{\textit{Uber Technologies, Inc., USA}}
\{djuloori, zplin, mattwill, eddy, smahajan\}@uber.com
}
\maketitle




\begin{abstract}
Maintaining a ``green" mainline branch—where all builds pass successfully—is crucial but challenging in fast-paced, large-scale software development environments, particularly with concurrent code changes in large monorepos. SubmitQueue, a system designed to address these challenges, speculatively executes builds and only lands changes with successful outcomes. However, despite its effectiveness, the system faces inefficiencies in resource utilization, leading to a high rate of premature build aborts and delays in landing smaller changes blocked by larger conflicting ones.

This paper introduces enhancements to SubmitQueue, focusing on optimizing resource usage and improving build prioritization. Central to this is our innovative probabilistic model, which distinguishes between changes with shorter and longer build times to prioritize builds for more efficient scheduling. By leveraging a machine learning model to predict build times and incorporating this into the probabilistic framework, we expedite the landing of smaller changes blocked by conflicting larger time-consuming changes. Additionally, introducing a concept of speculation threshold ensures that only the most likely builds are executed, reducing unnecessary resource consumption.

After implementing these enhancements across Uber's major monorepos  (Go, iOS, and Android), we observed a reduction in Continuous Integration (CI) resource usage by approximately 53\%, CPU usage by 44\%, and P95 waiting times by 37\%. These improvements highlight the enhanced efficiency of SubmitQueue in managing large-scale software changes while maintaining a green mainline.

\end{abstract}

\begin{IEEEkeywords}
Continuous Integration, Merge Queue, Monorepos, Build Time Prediction, Build Scheduling, Probabilistic Modeling, Speculative Execution, Version Control
\end{IEEEkeywords}

\section{Introduction}
\label{sec:introduction}

Hundreds of engineers frequently commit changes to a single repository in modern software development, particularly within large and fast-paced technology companies\cite{Potvin2016}. This scenario presents a significant challenge: efficiently managing these changes, quickly resolving conflicts, and ensuring that the mainline remains green. A mainline is considered green if all build steps—compilation, unit tests, and UI tests—are successfully executed for every commit point in the repository history. Maintaining a green mainline is critical for enabling rapid development and deployment cycles. However as highlighted in the study \cite{10.1145/3196398.3196421}, landing changes rapidly while keeping the mainline green becomes increasingly difficult as the codebase grows in size and complexity, further compounded by the concurrency of changes submitted by numerous developers.

While systems such as GitHub's Merge Queue \cite{Smythe2024}, GitLab's Merge Train \cite{Mishra2020}, LinkedIn's pre-merge validation \cite{Parikh2020}, and Airbnb's Evergreen \cite{Kudelka2022} and similar solutions \cite{Jain2023, trunk} aim to maintain a green mainline, they often fall short of rapidly landing changes while keeping the mainline green. These approaches lack either speculative execution or conflict resolution, leading to high resource usage and longer land times when earlier changes in the queue fail or during periods of high change velocity. 


SubmitQueue is designed to efficiently land changes while maintaining a green mainline, following the principles outlined in \cite{Ananthanarayanan2019}. At Uber, where over four thousand engineers work globally across thousands of microservices and tens of mobile apps, the system processes tens of thousands of changes each month. This scale involves managing hundreds of millions of lines of code across six major monorepos and seven programming languages. SubmitQueue facilitates hundreds of thousands of deployments and handles millions of configuration changes monthly. In this high-velocity development environment, ensuring the seamless and efficient integration of changes while landing them quickly into the mainline is crucial for maintaining service reliability, operational stability, and maximizing developer productivity.


SubmitQueue operates by speculating on the outcomes of all pending changes and constructing a speculation tree that outlines all possible builds for changes currently in the system. It uses a combination of a probabilistic model and a machine learning model to prioritize the builds most likely to succeed, executing them in parallel to minimize land times. This ensures that only changes passing all required checks are landed, thereby preserving the integrity of the mainline. Additionally, SubmitQueue performs conflict analysis between changes to prune the speculation tree, allowing independent changes to be built concurrently.

While SubmitQueue addresses many challenges in maintaining a green mainline, it still has certain limitations. As highlighted in the previous study \cite{Ananthanarayanan2019}, SubmitQueue has a strategy where it aborts ongoing builds of changes when the builds of the newly arrived changes are predicted to have a higher likelihood of success than those already in progress. This leads to two key issues:

\begin{enumerate}
    \item \textbf{High Resource Utilization}: A significant number of builds are prematurely aborted, with an estimated 40-65\% of builds being affected across major monorepos at Uber, which in turn leads to the need for scheduling additional builds for the changes whose builds were aborted.
    \item \textbf{Increased Waiting Times}: SubmitQueue processes changes in the order they are submitted. As a result, changes with shorter build times that arrive after a large, time-consuming change must wait for the larger change to either commit or reject before proceeding.  
\end{enumerate}

The introduction of Bypassing Large Diffs (BLRD) \cite{Lin2023} has partly addressed the issue of increased waiting times by introducing the concept of commutativity in change ordering. According to BLRD, if all the speculative builds for a smaller change, when blocked by a larger time-consuming conflicting change, have been evaluated and yield consistent outcomes, the smaller change can safely bypass the larger change and be landed or rejected based on the outcome. However, not all speculative builds for the smaller change are evaluated in most cases, as they are not prioritized. This occurs because SubmitQueue’s probabilistic model, as outlined in \cite{Ananthanarayanan2019}, assumes that only a single build is required to make a decision for each change and cannot distinguish between smaller and larger time-consuming changes to prioritize the builds accordingly. As a result, smaller changes continue to experience long waiting times when their speculative builds aren't prioritized, even though they could potentially bypass the larger changes ahead in the queue.

Addressing these limitations is crucial for two key reasons. First, resource utilization directly impacts operational costs, especially in fast-paced tech companies with high development velocity. Estimates suggest that large-scale CI systems, like those at Google and Mozilla, incur costs in the millions annually \cite{10.1145/3377811.3380437}. Inefficient build prioritization can further escalate these costs. For companies with limited budgets, this poses a significant barrier to adopting CI practices. Additionally, long waiting times reduce system efficiency and hinder developer productivity, as engineers face delays in landing their changes. 

This paper proposes a series of enhancements to the SubmitQueue system, focusing on optimizing resource usage and improving build prioritization to reduce waiting times. Specifically, we introduce a refined build prioritization strategy that leverages a machine learning model to predict build times and a novel probabilistic model to assess the eligibility of changes for bypassing to prioritize builds more effectively. Additionally, we present the concept of a speculation threshold to ensure that only the most probable builds are scheduled, further enhancing the system's efficiency.

Following the implementation of these enhancements across Uber’s major monorepos (Go, iOS, and Android), we saw notable improvements in key metrics: CI resource usage dropped by 53\%, weekly CPU hours by 44\%, and p95 waiting times by 37\%. These results highlight more efficient resource utilization, reduced consumption, and faster change landings, demonstrating the effectiveness of our strategy in optimizing SubmitQueue for large-scale CI environments.


The paper is structured as follows: We first discuss the background, highlighting the importance of maintaining a green mainline and the associated challenges. Next, we provide an overview of the SubmitQueue system, covering key concepts such as BLRD\cite{Lin2023}, BLRD eligibility, build completion estimation, and probabilistic build prioritization. We then explore the concept of speculation threshold and its impact on build scheduling. The paper continues with an in-depth discussion of the implementation and evaluation of these strategies, followed by conclusions and future work.

\section{Background}
\label{sec:background}
\subsection{\textbf{Importance of a Stable Mainline}}

As discussed in Section \ref{sec:introduction}, maintaining a green mainline—where all builds pass successfully is critical in large-scale, high-velocity development environments. A green mainline ensures stability, enabling seamless continuous integration, rapid development cycles, and consistent deployment of updates. 

As highlighted in \cite{Ananthanarayanan2019}, When the mainline turns ``red'' indicating a failure, it triggers a cascade of issues that can severely impact productivity, development timelines, and overall system stability. These issues include:

\begin{itemize}

\item \textbf{Delayed Rollouts:} A red mainline delays the deployment of new features and security patches, which can result in financial losses, especially in industries where time-to-market is critical. A study \cite{forsgren2018accelerate} shows that organizations with effective CI/CD practices deploy code 46 times more frequently, have 440 times faster commit-to-deploy times, and recover from downtime 170 times faster than low-performing organizations, underscoring the impact of delayed rollouts on revenue.


\item \textbf{Rollback Costs:} When failures occur, engineers must revert to the last stable state, often requiring complex cherry-picking and significant resource use. A study \cite{10.1145/3663529.3663856} on Atlassian projects (2021-2023) found that failed CI builds caused an average of 120 hours of wasted build time per project annually, resulting in substantial costs.

\item \textbf{Decreased Productivity}: A red mainline disrupts developer workflows by causing local build failures and forcing developers to work on code that may later be reverted. A study \cite{10.5555/2486788.2486846} has shown that process metrics like build stability are strong indicators of developer productivity. Unstable builds and frequent reverts significantly disrupt workflow, leading to wasted effort and frustration, as developers must spend time fixing issues that might not have existed in a stable mainline environment.
\end{itemize}

\subsection{\textbf{Other Approaches and Their Limitations}}



GitHub’s Merge Queue \cite{Smythe2024} and GitLab’s Merge Train \cite{Mishra2020} both operate by sequentially testing each change before merging. While they support parallel testing, these systems only speculate the success paths. When earlier changes in the queue fail, subsequent changes must be retested without the failed ones, which causes delays, especially when the failure rate is high. Additionally, both approaches lack conflict detection based on dependency graphs, leading to unnecessary test restarts when unrelated changes fail and causing further delays for subsequent changes.


\begin{figure*}[htbp] 
\centering 
\includegraphics[width=\textwidth]
{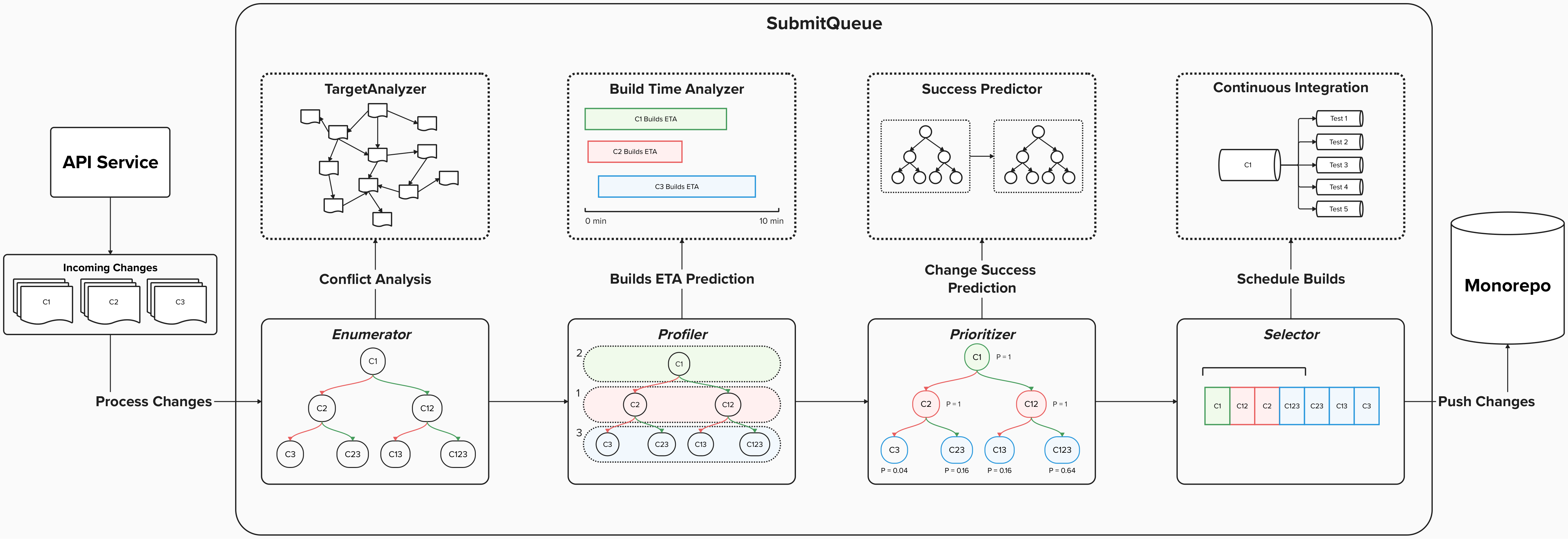} 
\caption{SubmitQueue Architecture} 
\label{fig:architecture} 
\end{figure*}

Airbnb’s Evergreen system \cite{Kudelka2022} addresses the serializability of changes in large monorepos by conducting conflict analysis and parallelizing the verification of independent changes. However, it falls short in scenarios with high-conflict periods or rapid change velocity. If pre-merge tests for earlier changes in the queue are lengthy, subsequent changes that have finished building must wait, limiting scalability in fast-paced environments. Additionally, if those pre-merge tests are prone to failures, it can result in repeated retesting of subsequent changes, leading to resource inefficiencies.


Aviator's Merge Queue \cite{Jain2023} is similar to SubmitQueue in that it speculatively executes builds of pending changes while performing conflict analysis. They introduced the concept of calculating a ``cutoff score'' to determine which speculation paths are worth building. While setting a cutoff score can reduce resource usage, it alone is insufficient. Under high-load conditions, smaller changes conflicting with larger changes must either wait for the larger changes to finish or face build delays if their scores fall below the cutoff, negatively impacting land times.

\subsection{\textbf{The Need for a Robust Change Management System}}
Given the limitations of other approaches, a more comprehensive solution is necessary to address the scale and concurrency of modern software development. Systems like SubmitQueue meet these demands by speculatively executing builds and landing only changes that pass all required checks, preventing mainline breakages before they occur.

By implementing systems like SubmitQueue, companies of any size can significantly enhance developer productivity, enabling faster release cycles while ensuring the highest levels of software quality. These systems help maintain a stable and reliable mainline, minimizing disruptions and reducing bottlenecks caused by concurrent changes. This paper builds on these foundations by introducing enhancements to optimize resource usage, reduce waiting times, and refine build prioritization for improved overall performance.

\section{System Overview}


When a change is submitted to SubmitQueue via the API service, it is added to a distributed queue for processing. The core service consists of several components responsible for executing all necessary build steps for each enqueued change. It ultimately determines whether to land or reject the change and the reason for rejection. Figure \ref{fig:architecture} illustrates the high-level architecture of SubmitQueue.


\subsection{\textbf{Enumerator}}

The Enumerator processes the queue of pending changes by constructing a speculation tree that outlines all possible builds for changes currently in the system. Using a target analyzer to identify potential conflicts between changes, the Enumerator (1) prunes unnecessary speculations to increase the likelihood of executing the remaining ones and (2) identifies independent changes that can be built in parallel, improving throughput.

\subsection{\textbf{Profiler}} 

The Profiler takes the speculation trees generated by the enumerator to create a profile for each tree, capturing information about the bypassing changes linked to each change within the tree. By predicting the build times of the nodes in the speculation tree, the build-time analyzer enables the Profiler to accurately identify the bypassing changes associated with each change.


\subsection{\textbf{Prioritizer}}

The Prioritizer calculates the probability of build needed for each node in the speculation tree by leveraging change bypassing data from the speculation tree profile and the success likelihood score of each change within SubmitQueue. It then ranks the builds based on these probabilities. The success likelihood score is predicted using a machine learning-powered success predictor, enabling the Prioritizer to make more informed build prioritization decisions.



\subsection{\textbf{Selector}} The Selector processes the prioritized builds and performs the following actions: (1) schedules high-probability builds for execution in the CI, (2) aborts ongoing builds that do not exist in the latest set of prioritized builds, and (3) safely commits changes to the monorepo once they meet all criteria for landing.
\section{Bypassing Large Diffs (BLRD)}
\label{sec:blrd}

\begin{figure}[htbp]
    \centering
    \begin{tikzpicture}
    \begin{axis}[
        width=\linewidth,
        xlabel={Month},
        ylabel={Conflict Rate (\%)},
        grid=major,
        ylabel near ticks,
        xlabel near ticks,
        xtick={1,2,3,4,5,6},
        xticklabels={Jan, Feb, Mar, Apr, May, Jun},
        legend style={at={(0.5,-0.2)}, anchor=north, legend columns=-1},
        ymin=0, ymax=80
    ]

    \addplot[color=darkred, mark=square*] coordinates {
        (1,30.04) (2,37.41) (3,41.13) (4,36.93) (5,41.17) (6,40.91)
    };
    \addlegendentry{Go}    
    
    \addplot[color=darkblue, mark=*] coordinates {
        (1,71.21) (2,66.57) (3,70.45) (4,60.78) (5,59.57) (6,69.45)
    };
    \addlegendentry{iOS}

    \addplot[color=darkgreen, mark=triangle*] coordinates {
        (1,62.54) (2,65.12) (3,72.33) (4,67.89) (5,64.37) (6,64.68)
    };
    \addlegendentry{Android}
    
    \end{axis}
    \end{tikzpicture}
    \caption{Monthly conflict rates across Go, iOS, and Android monorepos from January to June 2024.}
    \label{fig:monthly_conflict_rates}
\end{figure}
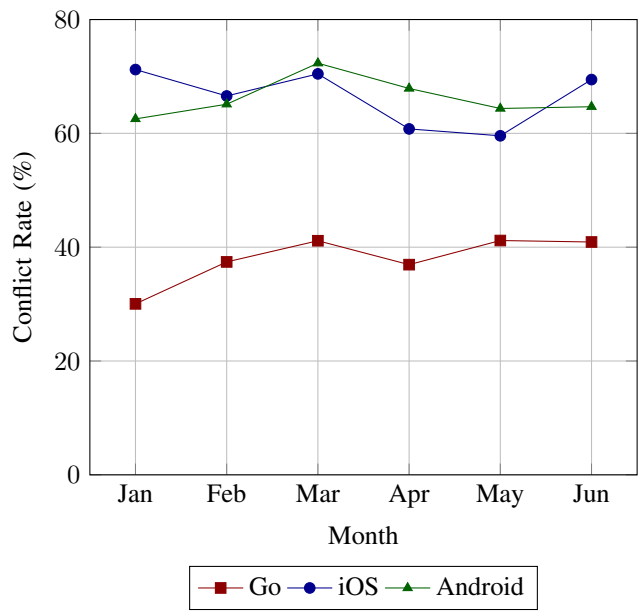

SubmitQueue executes builds in parallel to precompute results. However, it only decides whether to commit or reject a change once its corresponding build finishes \textit{and} reaches the head of the tree. As a result, smaller changes that arrive after larger conflicting changes are often delayed, even if their builds completed earlier. Large changes affecting many build targets can conflict with nearly every subsequent change processed by SubmitQueue. As illustrated in Figure \ref{fig:monthly_conflict_rates}, conflicts are frequent in the monorepos, and as the number of conflicts increases, the speculation tree grows deeper, further exacerbating delays.

As discussed in Section \ref{sec:introduction}, BLRD \cite{Lin2023} can expedite the landing of smaller changes if all of their speculative builds with the conflicting larger changes ahead yield the same outcome.

\renewcommand{\thefootnote}{\fnsymbol{footnote}}

\begin{figure}[htbp]
\centering
\includegraphics[width=8cm,height=8cm,keepaspectratio]{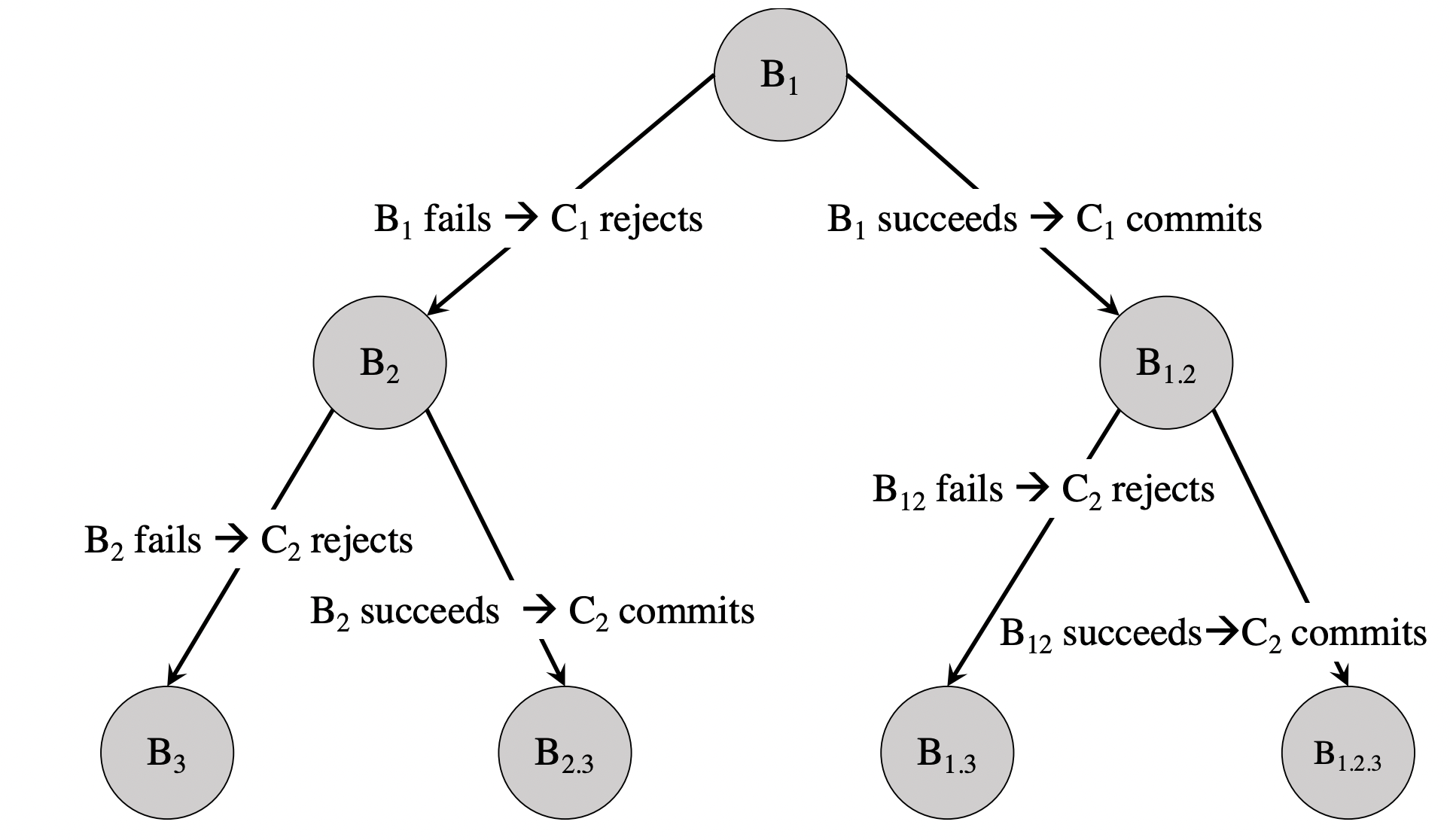}
\caption{Speculation tree of builds for conflicting changes \(C_1\), \(C_2\), and \(C_3\) that arrive in the mentioned order \cite{Ananthanarayanan2019}.\protect\footnotemark}
\label{fig:SpeculationTree}
\end{figure}

\footnotetext{Figure reproduced with authors' consent from the prior study \cite{Ananthanarayanan2019}.}

\renewcommand{\thefootnote}{\arabic{footnote}}


In Figure \ref{fig:SpeculationTree}, let \( H \) represent the current repository HEAD, and \( C_1 \), \( C_2 \), and \( C_3 \) be conflicting changes to be committed. Here, \( C_1 \) is a large time-consuming change, while \( C_2 \) is smaller and faster. For these changes, the following build steps are defined:
\begin{itemize}
    \item \( B_1 \) – Build steps for \( C_1 \) against \( H \).
    \item \( B_2 \) – Build steps for \( C_2 \) against \( H \).
    \item \( B_{1.2} \) – Build steps for \( C_2 \) against \( H + C_1 \).
    \item \( B_{1.2.3} \) – Build steps for \( C_3 \) against \( H + C_1 + C_2 \).
    \item \( B_{1.3} \) – Build steps for \( C_3 \) against \( H + C_1 \).
    \item \( B_{2.3} \) – Build steps for \( C_3 \) against \( H + C_2 \).
    \item \( B_3 \) – Build steps for \( C_3 \) against \( H \).
\end{itemize}

Let \( M(S, C) \) represent the state of the mainline after applying change \( C \) to state \( S \). SubmitQueue tests \( C_2 \) both on the current HEAD (\( B_2 \)) and against \( C_1 \) (\( B_{1.2} \)). If both speculative builds \( B_2 \) and \( B_{1.2} \) produce identical results, then the outcome of landing \( C_2 \) is independent of whether \( C_1 \) lands before or after. In this case, \( C_2 \) can be safely landed while \( C_1 \) is still in progress, ensuring:
\[
M(M(H, C_1), C_2) = M(M(H, C_2), C_1),
\]
thereby demonstrating that the order of landing \( C_1 \) and \( C_2 \) behaves \textbf{commutatively}. For more details and a complete proof of the BLRD concept, please refer to \cite{Lin2023}.

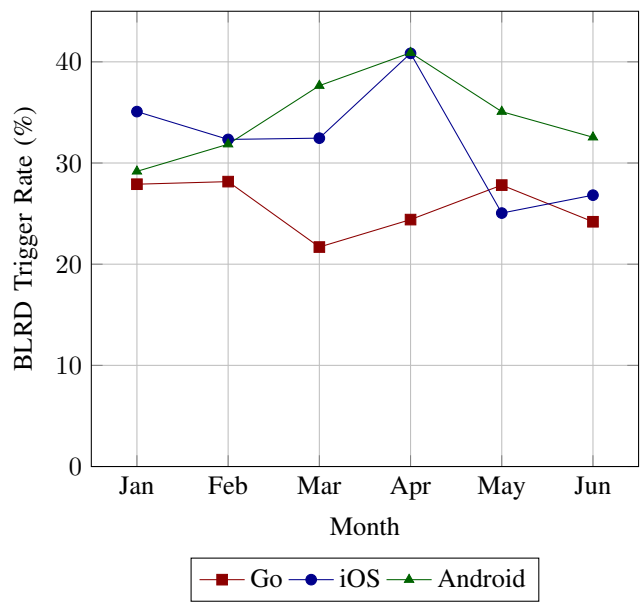
\begin{figure}[htbp]
    \centering
    \begin{tikzpicture}
    \begin{axis}[
        width=\linewidth,
        xlabel={Month},
        ylabel={BLRD Trigger Rate (\%)},
        grid=major,
        ylabel near ticks,
        xlabel near ticks,
        xtick={1,2,3,4,5,6},
        xticklabels={Jan, Feb, Mar, Apr, May, Jun},
        legend style={at={(0.5,-0.2)}, anchor=north, legend columns=-1},
        ymin=0, ymax=45
    ]

    \addplot[color=darkred, mark=square*] coordinates {
        (1,27.905738) (2,28.160644) (3,21.694029) (4,24.407646) (5,27.817221) (6,24.193006)
    };
    \addlegendentry{Go}

    \addplot[color=darkblue, mark=*] coordinates {
        (1,35.082459) (2,32.331556) (3,32.456861) (4,40.838988) (5,25.051653) (6,26.822392)
    };
    \addlegendentry{iOS}

    \addplot[color=darkgreen, mark=triangle*] coordinates {
        (1,29.174426) (2,31.857923) (3,37.644628) (4,40.895954) (5,35.078534) (6,32.545202)
    };
    \addlegendentry{Android}

    \end{axis}
    \end{tikzpicture}
    \caption{Monthly BLRD Trigger Rate for Go, Android, and iOS monorepos from January to June 2024.}
    \label{fig:monthly_blrd_trigger_rate}
\end{figure}

The BLRD trigger rate is calculated as the ratio of the total number of changes bypassed to the total number of changes that had to wait due to conflicting builds in progress. With rates ranging from 0\% to 45\% across all monorepos, this suggests significant room for improvement in optimizing the bypassing of smaller changes and reducing delays caused by the builds of larger conflicting changes ahead in the speculation tree.

\section{Probabilistic Model}
\label{sec:probabilitymodel}

SubmitQueue prioritizes builds based on their likelihood of being needed to optimize resource utilization. According to \cite{Ananthanarayanan2019}, the probability \( \displaystyle P^{\text{needed}}_{B_C} \) denotes the likelihood that the result of build \( B_C \) will be used to decide whether to commit or reject the change \( C \). The probabilistic model proposed in the prior research was based on the following assumptions:

\begin{enumerate}
    \item Only one build is necessary to determine the fate of a change.
    \item Changes are landed onto the mainline in the order they arrive in the queue.
\end{enumerate}

However, these assumptions no longer hold with the introduction of BLRD \cite{Lin2023}. As mentioned in Sections \ref{sec:introduction} and \ref{sec:blrd}, BLRD allows smaller changes to bypass larger, conflicting ones if all speculative builds of the smaller change are evaluated and produce consistent outcomes before the larger change completes. This requires SubmitQueue to evaluate multiple builds per change, ensuring smaller changes can bypass larger ones when eligible. The previous model's assumption of a single build determining a change's outcome is insufficient, as BLRD demands equal prioritization of all speculative builds to assess eligibility to bypass. However, evaluating all builds is expensive, because the number of builds grows exponentially with the depth of the speculation tree. Therefore, after the introduction of BLRD, the probabilistic model needs to prioritize builds to expedite the landing of the changes while optimizing resource usage.


\subsection{Enhanced Probabilistic Model}

The new model should focus on two key objectives:

\begin{enumerate}
    \item \textbf{Prioritize speculative builds for possible bypasses:} When changes can bypass larger conflicting changes ahead in the queue, all speculative builds of the eligible changes must be prioritized equally to allow smaller changes to land quickly.
    
    \item \textbf{Schedule builds for most likely paths in speculation tree for non-bypassing changes:} When a change's builds are likely to finish after the conflicting changes ahead in the queue are landed or rejected, only the most necessary speculative builds should be prioritized, as one build is sufficient to determine the outcome.
\end{enumerate}

Consider Figure \ref{fig:SpeculationTree}, which illustrates the new probabilistic model with multiple conflicting changes. The finish times for the changes are defined as follows:


\begin{itemize}
    \item The finish time of \( C_1 \) (\( FT_1 \)) is the time when build \( B_1 \) completes.
    \item The finish time of \( C_2 \) (\( FT_2 \)) is the time when both \( B_{1.2} \) and \( B_2 \) complete.
    \item The finish time of \( C_3 \) (\( FT_3 \)) is the time when all builds \( B_{1.2.3} \), \( B_{1.3} \), \( B_{2.3} \), and \( B_3 \) complete.
\end{itemize}

In this context, the expression \( P(FT_y < FT_x) \) represents the probability that the finish time of change \( C_y \) occurs before the finish time of change \( C_x \). These probabilities determine which builds are required, influencing the scheduling and prioritization of builds. This is because if \( C_y\)'s builds are unlikely to finish before \( C_x\)'s builds, there is little chance that BLRD will be used, and we only need to speculate \( C_y \) based on most likely outcome of \( C_x \). 

\begin{enumerate}
\item \textbf{\boldmath Case 1: \( FT_1 < FT_2 < FT_3 \)} \\
In this scenario, \( C_1 \) finishes before \( C_2 \), and \( C_2 \) finishes before \( C_3 \). Since there is no opportunity for bypassing, the root of the tree \( B_1 \) is always needed to determine the fate of \( C_1 \). The probabilities are:
\[
P^{\text{needed}}_{B_1} = 1, \quad P^{\text{needed}}_{B_{1.2}} = P^{\text{success}}_{B_1}, \quad P^{\text{needed}}_{B_2} = 1 - P^{\text{success}}_{B_1}
\]
\[
P^{\text{needed}}_{B_{1.2.3}} = P^{\text{success}}_{B_1} \times P^{\text{success}}_{B_2}
\]

This specific case was the primary focus of the probabilistic model presented in the prior study \cite{Ananthanarayanan2019}.

\item \textbf{\boldmath Case 2: \( FT_2 < FT_1 < FT_3 \)} \\
There is a chance that all builds of \( C_2 \) finish before \( C_1 \), allowing \( C_2 \) to bypass \( C_1 \), the outcomes of both builds \( B_2 \) and \( B_{1.2} \) are needed for determining whether \( C_2 \) lands before \( C_1 \). The probabilities are:
\[
P^{\text{needed}}_{B_2} = P^{\text{needed}}_{B_{1.2}} = P(FT_2 < FT_1)
\]

Since \( C_3 \) finishes last, its builds are dependent on the results of both \( C_1 \) and \( C_2 \).

\item \textbf{\boldmath Case 3: \( FT_1 < FT_3 < FT_2 \)} \\
There is also a possibility that \( C_1 \) finishes before \( C_3 \), and \( C_3 \) finishes before \( C_2 \). When BLRD is used for \( C_3 \), the build outcomes of \( C_2 \) do not affect the builds for \( C_3 \). The probabilities that we need different builds for \( C_3 \) to bypass \( C_2 \) are:
\[
P^{\text{needed}}_{B_{1.2.3}} = P^{\text{success}}_{B_1} \times P(FT_3 < FT_2)
\]
\[
P^{\text{needed}}_{B_{1.3}} = P^{\text{success}}_{B_1} \times P(FT_3 < FT_2)
\]
\[
P^{\text{needed}}_{B_{2.3}} = P^{\text{failure}}_{B_1} \times P(FT_3 < FT_2)
\]
\[
P^{\text{needed}}_{B_3} = P^{\text{failure}}_{B_1} \times P(FT_3 < FT_2)
\]

\item \textbf{\boldmath Case 4: \( FT_3 < FT_2 < FT_1 \)} \\
In this scenario, \( C_3 \) finishes before both \( C_2 \) and \( C_1 \). As a result, the build outcomes of \( C_1 \) and \( C_2 \) do not affect the builds for \( C_3 \), and similarly, the build outcome of \( C_1 \) do not affect the builds for \( C_2 \). The probabilities for \( C_3 \)’s builds are:
\[
P^{\text{needed}}_{B_{1.2.3}} = P(FT_3 < FT_1) \times P(FT_3 < FT_2)
\]
\[
P^{\text{needed}}_{B_{1.3}} = P(FT_3 < FT_1) \times P(FT_3 < FT_2)
\]
\[
P^{\text{needed}}_{B_{2.3}} = P(FT_3 < FT_1) \times P(FT_3 < FT_2)
\]
\[
P^{\text{needed}}_{B_3} = P(FT_3 < FT_1) \times P(FT_3 < FT_2)
\]

For \( C_2 \)’s builds:
\[
P^{\text{needed}}_{B_{1.2}} = P(FT_2 < FT_1), \quad P^{\text{needed}}_{B_2} = P(FT_2 < FT_1)
\]
\item \textbf{\boldmath Case 5: \( FT_3 < FT_1 < FT_2 \)} \\
In this scenario, \( C_3 \) finishes before \( C_1 \), and \( C_2 \) finishes last. Similar to Case 4, the builds for \( C_3 \) are unaffected by the build outcomes for \( C_1 \) and \( C_2 \), as \( C_3 \) completes first. However, unlike Case 4, \( C_2 \)'s builds depend on the outcome of \( C_1 \)'s builds, as \( C_1 \) finishes before \( C_2 \). As a result, the probabilities of the \( C_3 \) builds being needed remain the same as in Case 4, and the probabilities for \( C_2 \) builds remain the same as in Case 1.

\end{enumerate}

\subsection{Generalization}
The principles outlined in the cases above can be generalized for multiple changes in the queue. 


\[
P^{\text{needed}}_{B_C} = \prod_{C_i \in \mathcal{F}} P^{\text{outcome}}_{B_{C_i}} \times \prod_{C_j \in \mathcal{B}} P(FT_C < FT_{C_j})
\]



where:
\begin{itemize}
    \item \( P^{\text{needed}}_{B_C} \) represents the probability of a build being needed for a change \( C \).
    \item \( \mathcal{F} \) is the set of conflicting changes ahead of \( C \) in the queue that \( C \) does not bypass. 
    \item \( \mathcal{B} \) is the set of conflicting changes ahead that \( C \) may bypass.
    \item \( P^{\text{outcome}}_{B_{C_i}} \) represents the probability of the build outcome \( B_{C_i} \) for change \( C_i \) in \( \mathcal{F} \). The estimation of this probability is explained in the prior study \cite{Ananthanarayanan2019}.


    \item \( P(FT_C < FT_{C_j}) \) represents the probability that the finish time of \( C \) is less than the finish time of the change \( C_j \) in \( \mathcal{B} \), allowing \( C \) to bypass \( C_j \).
\end{itemize}

In extreme cases, when \( \prod_{C_j \in \mathcal{B}} P(FT_C < FT_{C_j}) \to 0 \), which can occur as the speculation depth increases in the tree, the system defaults to the original probabilistic model, prioritizing the most likely build needed for the change, i.e.,

\[
P^{\text{needed}}_{B_C} = \prod_{C_i \in \mathcal{A}} P^{\text{outcome}}_{B_{C_i}}
\]

where \( \mathcal{A} \) is the set of all changes ahead of \( C \) in the queue. This ensures efficient build prioritization and resource usage while minimizing unnecessary scheduling.

\section{Evaluating BLRD Eligibility}

BLRD may require an exponential number of builds to make a decision for a change, depending on its depth in the speculation tree, which can be computationally expensive. For example, in Figure 2, if the builds for \(C_2\) finish after those for \(C_1\), \(C_2\) cannot bypass \(C_1\) because \(C_1\) would have already been accepted or rejected. In such cases, only one of \(C_2\)'s builds is necessary. To improve cost efficiency, BLRD builds are scheduled only when there is a high probability that the builds of a change will finish before those of preceding conflicting changes. In Figure \ref{fig:SpeculationTree}, if \(P(FT_2 < FT_1)\) surpasses a predefined threshold, builds \(B_{1,2}\) and \(B_2\) for \(C_2\) are prioritized equally, optimizing resource allocation and minimizing unnecessary builds. This threshold, determined from empirical builds data, balances the need to prioritize smaller changes while reducing the risk of scheduling unneeded builds.





\section{Estimating the Probability of Build Completion Ordering}
\label{sec:buildcompletion}


Given two conflicting changes \( C_x \) and \( C_y \), arriving at times \( AT_x \) and \( AT_y \) (\( AT_x < AT_y \)), with respective finish times \( FT_x \) and \( FT_y \), and predicted build times \( T_x \) and \( T_y \), our goal is to estimate the probability that the build for \( C_y \) will finish before the build for \( C_x \), expressed as:

\[
P(FT_y < FT_x)
\]

The predicted build times \( T_x \) and \( T_y \) can be approximated as normally distributed random variables. We use the NGBoost model \cite{duan2020ngboost}, which is well-suited for probabilistic modeling, to predict build times. NGBoost learns patterns from historical builds to capture both the central tendency (mean) and variability (variance) in build times, smoothing out data irregularities and representing the predicted build times as a normal distribution. Thus,

\[
T_x \sim \mathcal{N}(\mu_x, \sigma_x^2), \quad T_y \sim \mathcal{N}(\mu_y, \sigma_y^2)
\]

where \( \mu_x \) and \( \mu_y \) represent the combined mean build times, and \( \sigma_x^2 \) and \( \sigma_y^2 \) represent the combined variances of the builds for changes \( C_x \) and \( C_y \). The combined mean and variance are computed by averaging the individual builds' means and variances. We seek to compute:

\[
P(FT_y < FT_x) = P((T_y + AT_y) < (T_x + AT_x))
\]

This expression can be rewritten as:

\[
P(T_y - T_x < AT_x - AT_y)
\]

We aim to compute the probability that the difference in build times, \( T_y - T_x \), is less than the difference in arrival times, \( AT_x - AT_y \). The random variables \( T_x \) and \( T_y \) are normally distributed, so the difference \( D = T_y - T_x \) is also normally distributed. The mean and variance of \( D \) are given by:

\[
\mu_D = \mu_y - \mu_x
\]
\[
\sigma_D^2 = \sigma_x^2 + \sigma_y^2
\]

Thus, \( D \sim \mathcal{N}(\mu_D, \sigma_D^2) \). The Z-score formula \cite{z_score} is used to standardize this difference:

\[
Z = \frac{(AT_x - AT_y) - (\mu_y - \mu_x)}{\sqrt{\sigma_x^2 + \sigma_y^2}}
\]

The Z-score measures how far the difference between arrival times is from the difference in build times, in standard deviations. Using the cumulative distribution function (CDF) of the standard normal distribution, the cumulative probability \( \Phi(Z) \) gives the likelihood that \( C_y \) finishes before \( C_x \):

\[
P(FT_y < FT_x) = \Phi(Z)
\]

where \( \Phi(Z) \) represents the value of the CDF for the Z-score.
\begin{enumerate}
    \item \textbf{Case 1: Close Arrival Times, Close Build Times}
    
    Request \( C_x \) arrives at 10:00 AM (\( AT_x = 0 \) minutes), with predicted build time \( \mu_x = 25 \) minutes and \( \sigma_x = 5 \) minutes. Request \( C_y \) arrives at 10:05 AM (\( AT_y = 5 \) minutes), with predicted build time \( \mu_y = 20 \) minutes and \( \sigma_y = 4 \) minutes. We compute the probability \( P(FT_y < FT_x) \) as follows:
    
    \[
    AT_x - AT_y = -5, \quad \mu_y - \mu_x = -5
    \]
    \[
    \sigma_D^2 = 5^2 + 4^2 = 41, \quad \sigma_D = \sqrt{41} \approx 6.40
    \]
    \[
    Z = \frac{-5 - (-5)}{6.40} = 0
    \]
    \( P(FT_y < FT_x) = \Phi(0) \approx 0.5\), meaning there is a 50\% probability that \( C_y \) finishes before \( C_x \).

    \item \textbf{Case 2: Large Difference in Arrival Times}
    
    Request \( C_x \) arrives at 10:00 AM (\( AT_x = 0 \) minutes), with predicted build time \( \mu_x = 20 \) minutes and \( \sigma_x = 4 \) minutes. Request \( C_y \) arrives at 6:00 PM (\( AT_y = 480 \) minutes), with predicted build time \( \mu_y = 5 \) minutes and \( \sigma_y = 2 \) minutes. 
    
    
    \[
    AT_x - AT_y = 0 - 480 = -480, \quad \mu_y - \mu_x = 5 - 20 = -15
    \]
    \[
    \sigma_D^2 = 4^2 + 2^2 = 16 + 4 = 20, \quad \sigma_D = \sqrt{20} \approx 4.47
    \]
    \[
    Z = \frac{-480 - (-15)}{4.47} = \frac{-465}{4.47} \approx -104.02
    \]
    Thus, \( P(FT_y < FT_x) = \Phi(-104.02) \approx 0\), meaning \( C_y \) is almost certain to finish after \( C_x \).

    \item \textbf{Case 3: Small Arrival Time Difference, Large Build Time Difference}
    
    Request \( C_x \) arrives at 10:00 AM (\( AT_x = 0 \) minutes), with predicted build time \( \mu_x = 35 \) minutes and \( \sigma_x = 6 \) minutes. Request \( C_y \) arrives at 10:01 AM (\( AT_y = 1 \) minute), with predicted build time \( \mu_y = 15 \) minutes and \( \sigma_y = 3 \) minutes. The probability \( P(FT_y < FT_x) \) is computed as follows:
    
    \[
    AT_x - AT_y = 0 - 1 = -1, \quad \mu_y - \mu_x = 15 - 35 = -20
    \]
    \[
    \sigma_D^2 = 6^2 + 3^2 = 36 + 9 = 45, \quad \sigma_D = \sqrt{45} \approx 6.71
    \]
    \[
    Z = \frac{-1 - (-20)}{6.71} = \frac{19}{6.71} \approx 2.83
    \]
    Thus, \( P(FT_y < FT_x) = \Phi(2.83) \approx 0.9977 \), meaning there is a 99.77\% probability that \( C_y \) finishes before \( C_x \).
    
\end{enumerate}

\section{Speculation Threshold}

The prior study \cite{Ananthanarayanan2019} did not address the implementation of a speculation threshold score for nodes in the speculation tree when selecting builds to schedule in CI. As a result, the system frequently scheduled more builds than necessary, including those with low \( P^{\text{needed}}_{B_C} \) scores, which led to increased build cancellations and significant resource wastage.

Setting a minimum threshold of \( P^{\text{needed}}_{B_C} \) can ensure that only the most probable nodes are selected for building. The threshold score should be informed by historical build data and adjusted based on the specific characteristics of the monorepos. However, setting the threshold too high can negatively impact land times, especially during high-load conditions. In such cases, more changes may be forced to wait for their builds, particularly if they conflict with larger changes ahead in the queue. Additionally, merely setting the speculation threshold could still leave smaller changes blocked by larger ones if the speculative builds of the smaller changes receive \( P^{\text{needed}}_{B_C}\) score lesser than the threshold.

By leveraging the probabilistic model suggested in Section \ref{sec:probabilitymodel}, the scores of speculative builds for smaller changes can be boosted if those changes are likely to bypass larger conflicting changes ahead. Thus, setting an appropriate speculation threshold, combined with the probabilistic model, strikes an optimal balance between resource usage and land times, ensuring efficient build scheduling without compromising throughput.

\section{Implementation}
\subsection{Core Service}
As outlined in the prior study \cite{Ananthanarayanan2019}, SubmitQueue is built as a robust Java service, leveraging MySQL \cite{MySQL} for backend storage, Bazel \cite{bazel} as a build system, Apache Helix \cite{ApacheHelix} for efficient sharding of queues across multiple machines, and RxJava \cite{rxjava3} for seamless event communication within processes. The introduction of BLRD \cite{Lin2023} and the incorporation of build-time prediction for build prioritization have fundamentally transformed SubmitQueue's core algorithm. Previously, the system employed a greedy, best-first, depth-wise traversal strategy, where nodes with the highest scores were visited first, and land/reject decisions for changes were based on the outcome of a single node. With BLRD, results from multiple nodes are often required to allow smaller changes to bypass larger, conflicting changes ahead in the speculation tree. To accommodate this, we have shifted to a level-order traversal approach, which enables simultaneous exploration of multiple speculative nodes for a given change, ensuring that nodes eligible for BLRD are prioritized equally. The earlier method, relying on the outcome of a single node, proved insufficient for managing the complexities introduced by BLRD.


\subsection{Model Training}
In our study, build refers to the compilation, testing, packaging, and publishing process triggered by code changes within a code repository. Each build encompasses multiple jobs that run concurrently to execute specific tasks, such as compiling different software components or executing various test suites. The parallel execution of jobs optimizes the overall build time, particularly for critical code changes that necessitate comprehensive testing.

As discussed in Section \ref{sec:buildcompletion}, We trained our models separately for each monorepo using the NGBoost (NGB) Regressor \cite{duan2020ngboost} to predict build times and handle uncertainties. These models were trained on Uber's native Machine Learning platform, Michelangelo \cite{wang2024predictive,hermann2017michelangelo}. The model outputs the mean and the standard deviation for each prediction. The mean provides the expected build duration, while the standard deviation offers a measure of uncertainty around this prediction. This dual-output approach enables us to manage prediction variability effectively and make more informed decisions based on the predicted build times and their associated uncertainties.


The dataset used to train the models comprises historical builds processed by SubmitQueue over the last three months, with the models being retrained weekly. We used the same feature set mentioned in the prior study \cite{Ananthanarayanan2019}, as these features are highly relevant and influential in predicting build times. Among all the features analyzed, the following have the highest influence on the model's predictions, listed in order of their importance:

\begin{itemize}
    \item \textbf{Targets Changed:} The number of targets modified in the change corresponding to the build.
    \item \textbf{Targets Added:} The number of new targets added in the change corresponding to the build.
    \item \textbf{Targets Removed:} The number of targets removed in the change corresponding to the build.
    \item \textbf{Conflicts Count:} The number of changes that the change corresponding to the build conflicts with.
    \item \textbf{Speculation Height:} The height of the node corresponding to the build in the speculation tree.
    \item \textbf{Added Lines:} The total number of new lines added in the change corresponding to the build.
    \item \textbf{Removed Lines:} The total number of lines removed in the change corresponding to the build.
    \item \textbf{ChangeSet Count:} The total number of files modified in the change corresponding to the build.
    \item \textbf{Commits Count:} The total number of commits in the change corresponding to the build.
    \item \textbf{Developer Name:} The author of the change corresponding to the build.
\end{itemize}

To train the model, the dataset was divided into training and validation sets, with 80\% of the data allocated for training and the remaining 20\% reserved for validation. This split ensures that the model is evaluated on unseen data, helping to assess its generalization capability and avoid overfitting.

To evaluate the performance of the NGB model, we used the Mean Absolute Percentage Error (MAPE) as the loss function. The MAPE is a metric that quantifies the average absolute errors as a percentage of the actual values, providing an intuitive understanding of the prediction error relative to the scale of the target variable. It is defined as:

\[
\text{MAPE} = \frac{1}{n} \sum_{i=1}^{n} \frac{\left| \hat{y}_i - y_i \right|}{y_i} \times 100\%
\]

where \( n \) is the number of predictions, \( \hat{y}_i \) is the predicted value, and \( y_i \) is the actual value.

Our best-performing models across different monorepos yielded the following MAPE values: 3\% for the Go monorepo, 6\% for the iOS monorepo, and 3.5\% for the Android monorepo. These results demonstrate that our model achieved high accuracy in predicting build times, with error rates generally falling below 5\%. The use of MAPE as our evaluation metric allowed us to assess the model's performance relative to the scale of the target values, ensuring that we could manage uncertainties effectively and make informed decisions based on the model’s predictions.

\subsection{Simulation System}

Testing new strategies in SubmitQueue can be costly and may result in longer land times, negatively impacting developer productivity. To mitigate these challenges, we developed a simulation system that replicates production-level traffic for a given monorepo. This system enables us to evaluate the effectiveness of algorithms by reconstructing requests and build contexts from a specific period. By simulating the order in which requests enter the queue and applying different algorithms, we can analyze key metrics such as land times, speculation penalties, waiting times for small changes, and total resource utilization. A key difference between the simulation and production environments is that speculative builds for a change may not always precisely replicate those in production, due to variations in traffic, system load, or other environmental factors. To address this, we estimate the build times of those builds by computing the mean of available build times for that change in production. The simulator is a crucial evaluation tool, allowing us to test and refine new algorithms before deploying them to production, ensuring that any changes are effective and cost-efficient.

\section{Evaluation}

SubmitQueue has been in production at Uber since 2018, following the strategy outlined in a previous study \cite{Ananthanarayanan2019}. On July 22nd, 2024, we rolled out a new strategy across three of Uber’s largest monorepos—Go, iOS, and Android—which consist of hundreds of millions of lines of code and handle thousands of daily changes submitted by hundreds of developers. To evaluate the effectiveness of this strategy, we tracked key performance metrics—weekly CPU hours, build-to-changes ratio, and P95 waiting times—over a 21-week period, from May 13, 2024, to September 30, 2024, capturing both pre- and post-rollout data. While our evaluation focused on Uber’s monorepos, the techniques presented here are language-agnostic and platform-independent. The following subsections summarize the performance improvements observed for each metric across the Go, iOS, and Android monorepos.

\subsection{Resource Usage}
\begin{figure}[htbp]
    \centering
    \begin{tikzpicture}
    \begin{axis}[
        width=\linewidth,
        xlabel={Week},
        ylabel={Builds-to-Changes Ratio},
        grid=major,
        ylabel near ticks,
        xlabel near ticks,
        legend style={at={(0.5,-0.2)}, anchor=north, legend columns=-1}
    ]
    
    \addplot[color=darkred, mark=square*] coordinates {
        (1,3.22)(2,4.01) (3,3.62) (4,3.80) (5,3.27) (6,3.70) (7,3.52) (8,2.81) (9,3.18) (10,4.04) (11,1.82) (12,1.69) (13,1.46) (14,1.70) (15,1.62) (16,1.70) (17,1.86) (18,1.42) (19,1.74) (20,1.76) (21,1.68)
    };
    \addlegendentry{Go}

    \addplot[color=darkblue, mark=*] coordinates {
        (1,3.69) (2,4.88) (3,3.85) (4,5.53) (5,5.40) (6,4.79) (7,5.58) (8,5.42) (9,4.84) (10,5.43) (11,3.77) (12,2.41) (13,2.34) (14,2.71) (15,2.60) (16,1.59) (17,1.20) (18,2.08) (19,2.38) (20,2.20) (21,1.98)
    };
    \addlegendentry{iOS}

    \addplot[color=darkgreen, mark=triangle*] coordinates {
        (1,5.53) (2,5.71) (3,5.34) (4,6.00) (5,5.07) (6,5.62) (7,5.41) (8,4.44) (9,5.87) (10,6.26) (11,2.86) (12,1.70) (13,1.60) (14,1.76) (15,1.72) (16,1.40) (17,1.30) (18,1.44) (19,1.34) (20,1.42) (21,1.48)
    };
    \addlegendentry{Android}

    \draw[densely dotted, thick] (axis cs:10.5,\pgfkeysvalueof{/pgfplots/ymin}) -- (axis cs:10.5,\pgfkeysvalueof{/pgfplots/ymax});
    \addlegendimage{densely dotted, thick, line legend}
    \addlegendentry{Rollout}
    
    \end{axis}
    \end{tikzpicture}
    \caption{Weekly trend of Builds-to-Changes ratio across Go, iOS, and Android monorepos.}
    \label{fig:multi_monorepo_build_to_change_ratio}
\end{figure}
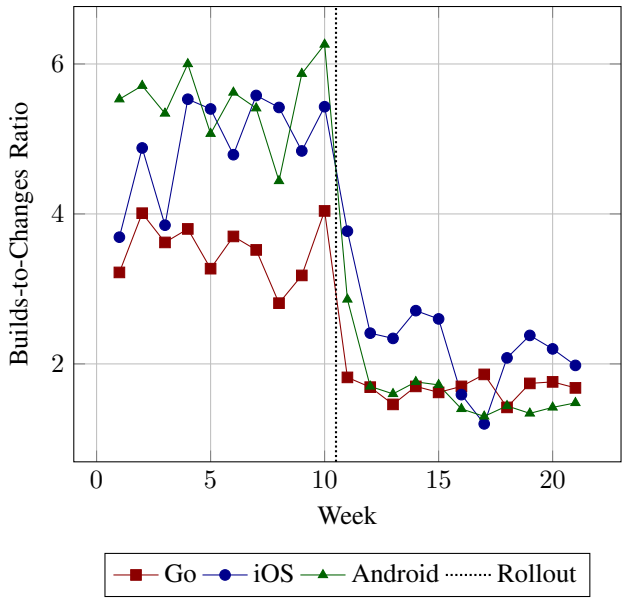

The build-to-changes ratio is a critical metric for assessing resource efficiency in our CI pipeline. Figure \ref{fig:multi_monorepo_build_to_change_ratio} illustrates the ratio trends over the 21-week evaluation period.

In the first 10 weeks, prior to the rollout, the ratio fluctuated between 3 and 6, with iOS and Android showing greater variability. The Android monorepo peaked at a ratio of 6 in Week 7, highlighting inefficiencies in resource utilization due to excessive builds per change.

After the rollout, a sharp decline in the ratio was observed across all monorepos. The Go monorepo saw a reduction of \textbf{45.45\%} (from a pre-rollout average of 3.39 to a post-rollout average of 1.85), iOS decreased by \textbf{47.86\%} (from 4.93 to 2.57), and Android achieved the largest reduction of \textbf{64.02\%}, dropping from 5.43 to 1.96. These results demonstrate improved resource allocation and more efficient build scheduling post-rollout.

\subsection{CPU Hours Consumption}
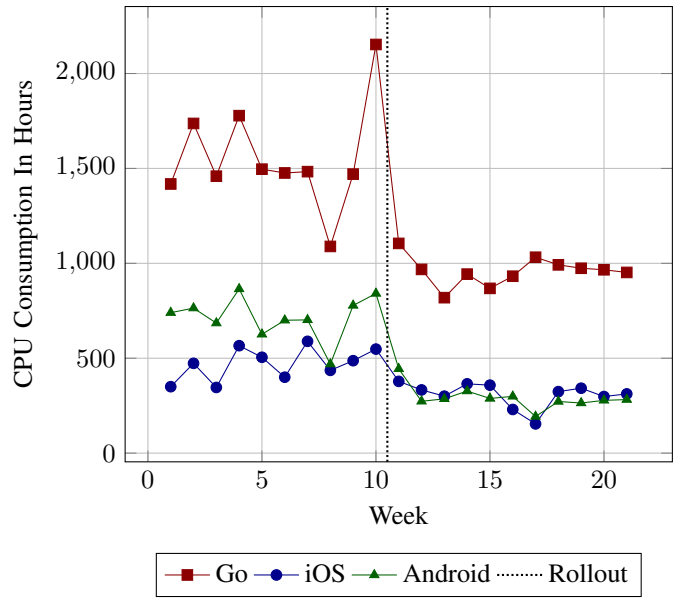
\begin{figure}[htbp]
    \centering
    \begin{tikzpicture}
    \begin{axis}[
        width=\linewidth,
        xlabel={Week},
        ylabel={CPU Consumption In Hours},
        grid=major,
        ylabel near ticks,
        xlabel near ticks,
        legend style={at={(0.5,-0.2)}, anchor=north, legend columns=-1}
    ]
 
    \addplot[color=darkred, mark=square*] coordinates {
        (1,1418) (2,1737) (3,1459) (4,1778) (5,1496) (6,1476) (7,1483) (8,1089) (9,1470) (10,2153) (11,1105) (12,968) (13,819) (14,943) (15,868) (16,932) (17,1032) (18,992) (19,974) (20,966) (21,952) 
    };
    \addlegendentry{Go}
   
    \addplot[color=darkblue, mark=*] coordinates {
        (1,350) (2,473) (3,346) (4,566) (5,505) (6,400) (7,589) (8,436) (9,487) (10,548) (11,378) (12,333) (13,300) (14,365) (15,358) (16,230) (17,154) (18,324) (19,342) (20,298) (21,312)
    };
    \addlegendentry{iOS}
     
    \addplot[color=darkgreen, mark=triangle*] coordinates {
        (1,740) (2,764) (3,685) (4,865) (5,626) (6,700) (7,702) (8,469) (9,778) (10,841) (11,444) (12,273) (13,286) (14,327) (15,288) (16,299) (17,193) (18,272) (19,264) (20,278) (21,282)
    };
    \addlegendentry{Android}

    \draw[densely dotted, thick] (axis cs:10.5,\pgfkeysvalueof{/pgfplots/ymin}) -- (axis cs:10.5,\pgfkeysvalueof{/pgfplots/ymax});
    \addlegendimage{densely dotted, thick, line legend}
    \addlegendentry{Rollout}
    
    \end{axis}
    \end{tikzpicture}
    \caption{Weekly CPU hours consumption across Go, iOS, and Android monorepos.}
    \label{fig:multi_monorepo_cpu_savings}
\end{figure}

Figure \ref{fig:multi_monorepo_cpu_savings} shows the weekly CPU hours consumed across Go, iOS, and Android monorepos during the evaluation period.

Before the rollout, CPU usage was consistently high, particularly in the Go monorepo, which peaked at around 2,000 hours in Week 8. Android and iOS fluctuated between 500 and 800 hours and 400 to 600 hours, respectively.

Following the rollout, CPU hours dropped significantly across all monorepos. Go's CPU consumption fell by \textbf{44.70\%} (from a pre-rollout average of 1,485 to a post-rollout average of 821 hours), iOS decreased by \textbf{34.86\%} (from 472 to 307 hours), and Android saw a reduction of \textbf{52.23\%} (from 729 to 348 hours). These reductions highlight the efficiency gains from minimizing unnecessary builds, leading to substantial cost savings and improved scalability.

\subsection{P95 Waiting Times}

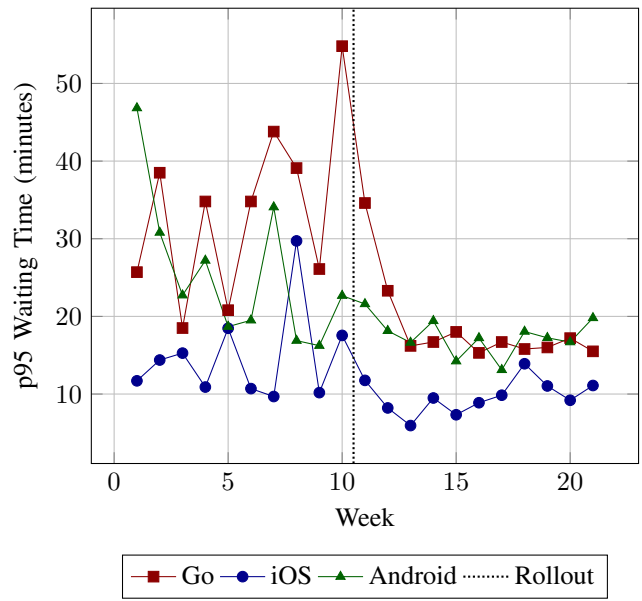
\begin{figure}[htbp]
    \centering
    \begin{tikzpicture}
    \begin{axis}[
        width=\linewidth,
        xlabel={Week},
        ylabel={p95 Waiting Time (minutes)},
        grid=major,
        ylabel near ticks,
        xlabel near ticks,
        legend style={at={(0.5,-0.2)}, anchor=north, legend columns=-1}
    ]  
    \addplot[color=darkred, mark=square*] coordinates {
        (1,25.7) (2,38.5) (3,18.5) (4,34.8) (5,20.8) (6,34.8) (7,43.8) (8,39.1) (9,26.1) (10,54.8) (11,34.6) (12,23.3) (13,16.2) (14,16.7) (15,18.0) (16,15.3) (17,16.7) (18,15.8) (19,16.0) (20, 17.2) (21, 15.5)
    };
    \addlegendentry{Go}
   
    \addplot[color=darkblue, mark=*] coordinates {
        (1,11.70) (2,14.38) (3,15.27) (4,10.90) (5,18.47) (6,10.70) (7,9.68) (8,29.72) (9,10.18) (10,17.55) (11,11.76) (12,8.21) (13,5.94) (14,9.49) (15,7.33) (16,8.88) (17,9.85) (18,13.90) (19,11.04) (20,9.20) (21,11.10)
    };
    \addlegendentry{iOS}

    \addplot[color=darkgreen, mark=triangle*] coordinates {
        (1,46.83) (2,30.80) (3,22.72) (4,27.20) (5,18.68) (6,19.50) (7,34.07) (8,16.88) (9,16.23) (10,22.67) (11,21.60) (12,18.14) (13,16.63) (14,19.42) (15,14.23) (16,17.22) (17,13.11) (18,18.05) (19,17.24) (20,16.69)(21,19.80)
    };
    \addlegendentry{Android}

    \draw[densely dotted, thick] (axis cs:10.5,\pgfkeysvalueof{/pgfplots/ymin}) -- (axis cs:10.5,\pgfkeysvalueof{/pgfplots/ymax});
    \addlegendimage{densely dotted, thick, line legend}
    \addlegendentry{Rollout}
    
    \end{axis}
    \end{tikzpicture}
    \caption{Weekly p95 Waiting Times for changes in Go, iOS, and Android monorepos.}
    \label{fig:multi_monorepo_waiting_times}
\end{figure}

CI resource usage is not the only metric to optimize in this work. Trivially, we can set the speculation threshold so high that SubmitQueue rarely speculates more than one path, which would reduce CI resource usage too. However, that would also reduce the efficiency of SubmitQueue, because BLRD, which requires multiple speculation paths, would not be used. 
When the conditions for BLRD is not met, a change has to wait in SubmitQueue even though all its speculative builds finish. We monitor P95 waiting time before and after the rollout to make sure the reduction of resource usage does not reduce the efficiency. Figure \ref{fig:multi_monorepo_waiting_times} illustrates the weekly P95 waiting times across the Go, iOS, and Android monorepos.

Initially, waiting times fluctuated significantly, with the Go monorepo showing peaks around weeks 5 and 10. After the new strategy was implemented in Week 11, all monorepos showed stabilization, particularly in iOS and Android, where waiting times consistently dropped to lower levels than pre-rollout. Go also demonstrated reduced variability, with more stable waiting times after Week 15.

Overall, the P95 waiting times were reduced by \textbf{44.67\%} for Go (from a pre-rollout average of 33.69 minutes to a post-rollout average of 18.64 minutes), \textbf{33.32\%} for iOS (from 14.86 minutes to 9.91 minutes), and \textbf{31.66\%} for Android (from 25.36 minutes to 17.33 minutes). This reduction signifies the impact of the new build prioritization strategy in expediting smaller changes landing using BLRD \cite{Lin2023}.

\section{Related Work}

\subsection{Prediction Systems in CI}

BuildFast \cite{9286064} and SmartBuildSkip \cite{10.1145/3377811.3380437} introduced systems for predicting build outcomes, while DL-CIBuild \cite{10.1007/s10515-021-00319-5} used LSTM models to predict CI build failures, outperforming traditional methods. These approaches focus on predicting build outcomes rather than estimating when builds will be complete—a crucial factor in CI scheduling.

A study \cite{Bisong_2017} predicted build times using the TravisTorrent dataset, with Random Forest \cite{breiman2001random} showing strong performance. We extended this by using NGBoost \cite{duan2020ngboost}, which predicts both build times and their uncertainties, essential for handling variability in real-world CI. Our system integrates these predictions into an adaptive scheduling framework, optimizing resource allocation and reducing wait times, especially for smaller changes bypassing larger ones.

\subsection{CI Scheduling Algorithms}

The study on scheduling algorithms for improved CI performance \cite{nilsson2023scheduling} demonstrated the advantages of leveraging expected processing times to improve scheduling decisions in CI environments. This study found that Shortest Processing Time (SPT) and Gupta’s algorithm \cite{DBLP:journals/corr/GuptaMUX17,10.1287/moor.2021.1149} significantly reduce wait times and improve system performance. Similarly, our use of NGBoost\cite{duan2020ngboost} not only predicts build times but also incorporates uncertainties, making it comparable to Gupta’s stochastic scheduling approach, which optimizes job uncertainty in dynamic environments. Both approaches highlight the importance of data-driven predictions in improving the efficiency of CI systems.

\subsection{Commutativity in Concurrent Systems}

As demonstrated in BLRD \cite{Lin2023}, the concept of commutativity in build operations parallels broader research on concurrent systems. A method for analyzing commutativity in concurrent programs using state-chart graph representations is presented in \cite{debnath2019analysiscommutativitystatechartgraph}. This work provides a formal framework for understanding when operations in a concurrent system can be safely reordered, which is conceptually similar to the commutativity property exploited by BLRD in the context of build systems. 

\section{Limitations and Future Work}

While our approach has significantly improved SubmitQueue, several areas for further exploration and enhancements exist. Future work could focus on:


\textbf{Dynamic Speculation Thresholds:} Dynamically adjusting scheduling thresholds for better resource management in speculative execution has been explored in various systems. For example, certain strategies focus on improving resource allocation by detecting task inefficiencies in real-time and minimizing unnecessary speculative tasks \cite{cmc.2020.04604}. Similar adaptive scheduling techniques could benefit SubmitQueue, which currently uses static thresholds that may not align with real-time workload demands. Future work will explore dynamic strategies that adjust speculation thresholds based on the system's current state to prevent inefficiencies.


\textbf{Change Batching:} While BLRD \cite{Lin2023} addresses the issue of post-build evaluation waiting, changes often face delays to begin build scheduling due to resource constraints. Recent research has explored various batching techniques to optimize this process. DynamicBatching \cite{10.1145/3611643.3616255} introduced a technique that adapts batch size based on the request traffic. Another approach \cite{9609234} includes using weighted historical failure rates and mining historical test failures to inform batching decisions.  SubmitQueue's ability to predict build times and the likelihood of success can be further leveraged in accordance with these techniques. We can improve request throughput and optimize resource utilization by batching changes with a high probability of success and similar build durations. 

\section{Conclusion}\label{sec:Conclusions}

In this paper, we introduced enhancements to SubmitQueue focused on optimizing resource usage and reducing waiting times for changes in large-scale development environments. While BLRD \cite{Lin2023} partially addressed prolonged waiting times for changes with shorter build times, as discussed in \cite{Ananthanarayanan2019}, Our approach further enhances this by leveraging machine learning for build-time predictions and incorporating it into our novel probabilistic framework. These improvements tackle inefficiencies in resource consumption and delays caused by larger conflicting changes. Additionally, introducing a speculation threshold ensures that only the most probable builds are scheduled, further streamlining the process. These advancements position SubmitQueue as a highly efficient tool for managing software changes, adaptable to environments ranging from small teams to large enterprises. Adopting systems like SubmitQueue can lead to faster release cycles, reduced costs, and higher software quality, fostering more efficient industry-wide development practices. Future work will explore further optimization opportunities.

\section*{Acknowledgment}
The authors sincerely thank Chris Zhang and Akshay Utture from Uber's Programming Systems Group for their valuable suggestions. We also appreciate the support of Shesh Patel, Matt Morgan, and Anshu Chadha, leaders of Uber's Developer Platform, in bringing this project to fruition.


\bibliographystyle{unsrt}
\bibliography{refs}

\begin{thebibliography}{10}

\bibitem{Potvin2016}
Rachel Potvin and Josh Levenberg.
\newblock Why google stores billions of lines of code in a single repository.
\newblock {\em Communications of the ACM}, 59:78--87, 2016.

\bibitem{10.1145/3196398.3196421}
Jo\~{a}o~Helis Bernardo, Daniel~Alencar da~Costa, and Uir\'{a} Kulesza.
\newblock Studying the impact of adopting continuous integration on the delivery time of pull requests.
\newblock In {\em Proceedings of the 15th International Conference on Mining Software Repositories}, MSR '18, page 131–141, New York, NY, USA, 2018. Association for Computing Machinery.

\bibitem{Smythe2024}
Will Smythe and Lawrence Gripper.
\newblock How github uses merge queue to ship hundreds of changes every day.
\newblock \url{https://github.blog/engineering/engineering-principles/how-github-uses-merge-queue-to-ship-hundreds-of-changes-every-day/}, March 2024.

\bibitem{Mishra2020}
Veethika Mishra.
\newblock How to use merge train pipelines with gitlab.
\newblock \url{https://about.gitlab.com/blog/2020/12/14/merge-trains-explained/}, December 2020.

\bibitem{Parikh2020}
Niket Parikh.
\newblock How linkedin handles merging code in high-velocity repositories.
\newblock \url{https://www.linkedin.com/blog/engineering/optimization/continuous-integration/}, April 2020.

\bibitem{Kudelka2022}
Janusz Kudelka and Joel Snyder.
\newblock Evergreen: Building airbnb’s merge queue with kafka streams.
\newblock \url{https://www.confluent.io/events/kafka-summit-london-2022/evergreen-building-airbnbs-merge-queue-with-kafka-streams/}, April 2022.

\bibitem{Jain2023}
Ankit Jain.
\newblock Merge strategies to keep builds healthy at scale.
\newblock \url{https://www.aviator.co/blog/merge-strategies-at-scale/}, September 2023.

\bibitem{trunk}
{Trunk}.
\newblock Trunk - the fast lane for your prs.
\newblock \url{https://trunk.io/}, 2024.

\bibitem{Ananthanarayanan2019}
Sundaram Ananthanarayanan, Masoud~Saeida Ardekani, Denis Haenikel, Balaji Varadarajan, Simon Soriano, Dhaval Patel, and Ali-Reza Adl-Tabatabai.
\newblock Keeping master green at scale.
\newblock In {\em Proceedings of the Fourteenth EuroSys Conference 2019}, New York, NY, USA, 2019. Association for Computing Machinery.

\bibitem{Lin2023}
Zhongpeng Lin and Matthew Williams.
\newblock Bypassing large diffs in submitqueue.
\newblock \url{https://www.uber.com/blog/bypassing-large-diffs-in-submitqueue/}, August 2023.

\bibitem{10.1145/3377811.3380437}
Xianhao Jin and Francisco Servant.
\newblock A cost-efficient approach to building in continuous integration.
\newblock In {\em Proceedings of the ACM/IEEE 42nd International Conference on Software Engineering}, ICSE '20, page 13–25, New York, NY, USA, 2020. Association for Computing Machinery.

\bibitem{forsgren2018accelerate}
N.~Forsgren, J.~Humble, and G.~Kim.
\newblock {\em Accelerate: The Science of Lean Software and DevOps: Building and Scaling High Performing Technology Organizations}.
\newblock IT Revolution Press, 2018.

\bibitem{10.1145/3663529.3663856}
Yang Hong, Chakkrit Tantithamthavorn, Jirat Pasuksmit, Patanamon Thongtanunam, Arik Friedman, Xing Zhao, and Anton Krasikov.
\newblock Practitioners' challenges and perceptions of ci build failure predictions at atlassian, 06 2024.

\bibitem{10.5555/2486788.2486846}
Foyzur Rahman and Premkumar Devanbu.
\newblock How, and why, process metrics are better.
\newblock In {\em Proceedings of the 2013 International Conference on Software Engineering}, ICSE '13, page 432–441. IEEE Press, 2013.

\bibitem{duan2020ngboost}
Tony Duan, Anand Avati, Daisy~Yi Ding, Khanh~K. Thai, Sanjay Basu, Andrew~Y. Ng, and Alejandro Schuler.
\newblock Ngboost: Natural gradient boosting for probabilistic prediction.
\newblock In {\em Proceedings of the 37th International Conference on Machine Learning}, ICML '20, page 2690–2700. JMLR.org, 2020.

\bibitem{z_score}
Stephanie Glen.
\newblock Z-score: Definition, formula and calculation, 2024.

\bibitem{MySQL}
{MySQL AB}.
\newblock Mysql: The world's most popular open source database, 2024.

\bibitem{bazel}
{Google}.
\newblock Bazel - a fast, scalable, multi-language and extensible build system.
\newblock \url{https://bazel.build/}, 2024.

\bibitem{ApacheHelix}
{Apache Software Foundation}.
\newblock Apache helix, 2024.

\bibitem{rxjava3}
{ReactiveX}.
\newblock Rxjava: Reactive extensions for the jvm.
\newblock \url{https://github.com/ReactiveX/RxJava}, 2022.

\bibitem{wang2024predictive}
Kai Wang, Mingshi Cai, Jingya Wang, and Eric Chen.
\newblock From predictive to generative - how michelangelo accelerates uber's ai journey.
\newblock \url{https://www.uber.com/blog/from-predictive-to-generative-ai/}, 5 2024.

\bibitem{hermann2017michelangelo}
Jeremy Hermann and Mike Del~Balso.
\newblock Meet michelangelo: Uber's machine learning platform.
\newblock \url{https://www.uber.com/blog/michelangelo-machine-learning-platform/}, 2017.
\newblock Accessed: October 4, 2024.

\bibitem{9286064}
Bihuan Chen, Linlin Chen, Chen Zhang, and Xin Peng.
\newblock Buildfast: History-aware build outcome prediction for fast feedback and reduced cost in continuous integration.
\newblock In {\em 2020 35th IEEE/ACM International Conference on Automated Software Engineering (ASE)}, pages 42--53, 2020.

\bibitem{10.1007/s10515-021-00319-5}
Islem Saidani, Ali Ouni, and Mohamed~Wiem Mkaouer.
\newblock Improving the prediction of continuous integration build failures using deep learning.
\newblock {\em Automated Software Engg.}, 29(1), May 2022.

\bibitem{Bisong_2017}
Ekaba Bisong, Eric Tran, and Olga Baysal.
\newblock Built to last or built too fast? evaluating prediction models for build times.
\newblock In {\em 2017 IEEE/ACM 14th International Conference on Mining Software Repositories (MSR)}. IEEE, May 2017.

\bibitem{breiman2001random}
Leo Breiman.
\newblock Random forests.
\newblock {\em Machine Learning}, 45(1):5--32, 2001.

\bibitem{nilsson2023scheduling}
Zacharias~Faleberg Nilsson and Freddy Abrahamsson.
\newblock Scheduling algorithms for improved ci performance.
\newblock Master's thesis, Chalmers University of Technology and University of Gothenburg, Gothenburg, Sweden, June 2023.
\newblock Department of Computer Science and Engineering.

\bibitem{DBLP:journals/corr/GuptaMUX17}
Varun Gupta, Benjamin Moseley, Marc Uetz, and Qiaomin Xie.
\newblock Stochastic online scheduling on unrelated machines.
\newblock {\em CoRR}, abs/1703.01634, 2017.

\bibitem{10.1287/moor.2021.1149}
Varun Gupta, Benjamin Moseley, Marc Uetz, and Qiaomin Xie.
\newblock Corrigendum: Greed works—online algorithms for unrelated machine stochastic scheduling.
\newblock {\em Math. Oper. Res.}, 46(3):1230–1234, August 2021.

\bibitem{debnath2019analysiscommutativitystatechartgraph}
Kishore Debnath, Christina Peterson, and Damian Dechev.
\newblock Analysis of commutativity with state-chart graph representation of concurrent programs, 2019.

\bibitem{cmc.2020.04604}
Yinghang Jiang, Qi~Liu, Williams Dannah, Dandan Jin, Xiaodong Liu, and Mingxu Sun.
\newblock An optimized resource scheduling strategy for hadoop speculative execution based on non-cooperative game schemes, 2020.

\bibitem{10.1145/3611643.3616255}
Emad Fallahzadeh, Amir~Hossein Bavand, and Peter~C. Rigby.
\newblock Accelerating continuous integration with parallel batch testing.
\newblock In {\em Proceedings of the 31st ACM Joint European Software Engineering Conference and Symposium on the Foundations of Software Engineering}, ESEC/FSE 2023, page 55–67, New York, NY, USA, 2023. Association for Computing Machinery.

\bibitem{9609234}
Amir~Hossein Bavand and Peter~C. Rigby.
\newblock Mining historical test failures to dynamically batch tests to save ci resources.
\newblock In {\em 2021 IEEE International Conference on Software Maintenance and Evolution (ICSME)}, pages 217--226, 2021.

\end{thebibliography}
\end{document}